\documentclass[twocolumn,aps,showpacs,preprintnumbers,amsmath,amssymb,superscriptaddress]{revtex4-1}
\usepackage[pdftex]{graphicx}
\usepackage{amsmath}
\usepackage{amsfonts}
\usepackage{amssymb}
\usepackage{color}
\usepackage{float}
\usepackage{color}
\usepackage{natbib}
\usepackage{appendix}
\begin{document}
\title{Optimal Single Quantum Dot Heat-to-pure-spin-current Converters}
\bigskip
\author{Siddharth Buddhiraju}
\author{Bhaskaran Muralidharan}
\affiliation{Department of Electrical Engineering, Indian Institute of Technology Bombay, Powai, Mumbai-400076, India}
\date{\today}
\medskip
\widetext
\begin{abstract}
We delve into the conditions under which a quantum dot thermoelectric setup may be tuned to realize an optimal heat-to-pure-spin-current converter. It is well known that a heat-to-pure-spin-current converter may be realized using a non-interacting quantum dot with a spin-split energy spectrum under particle hole symmetry conditions. However, with the inclusion of Coulomb interaction $U$, ubiquitous in typical quantum dot systems, the relevant transport physics is expected to be altered. In this work, we provide a detailed picture of thermoelectric pure spin currents at various Coulomb interaction parameters $U$ and describe the conditions necessary for an exact cancellation of charge transport between energy levels $\epsilon$ and their Coulomb-charged partner levels $\epsilon+U$, so as to yield the largest terminal pure spin currents. A non-trivial aspect pointed out here is that at sufficiently large values of $U$ ($\ge U_0$), pure spin currents tend to optimize at points other than where the particle-hole symmetry occurs. It is also ascertained that a global maximum of pure spin current is generated at a typical value of the interaction parameter $U$. These optimum conditions may be easily realized using a typical gated quantum dot thermoelectric transport setup.
\end{abstract}
\maketitle
\section{Introduction}
Spin currents form an important aspect of energy efficient spintronic information processing \cite{Bader}. Spin currents generated electrically, optically \cite{Bader} or via thermoelectric means \cite{Bauer_2011} find direct applications in spin transfer torque based magnetic switching \cite{slon,berger,kat,ralph1,Bauer_TSTT,Zhang_1,Brataas_2}, spin torque oscillators \cite{Ralph_1,Ralph_2,Grollier} and domain wall manipulation \cite{Bader,Parkin}, to name a few. Pure spin currents, or spin currents without accompanying charge currents \cite{Pure_spin_2,pure_spin_1,Pure_spin_3,Pure_spin_4,Spin_current_1} are important specifically for low power applications, allowing in some cases the possibility of dissipationless spin transport \cite{Spin_current_1,Murakami,Sloncewski}. The efficient thermoelectric generation of pure spin currents across low-dimensional systems is thus an important aspect of spintronic energy harvesting \cite{Bauer_2011,vanWees}. \\
\begin{figure}
\centering
	\includegraphics[width=2.85in, height = 2.3in]{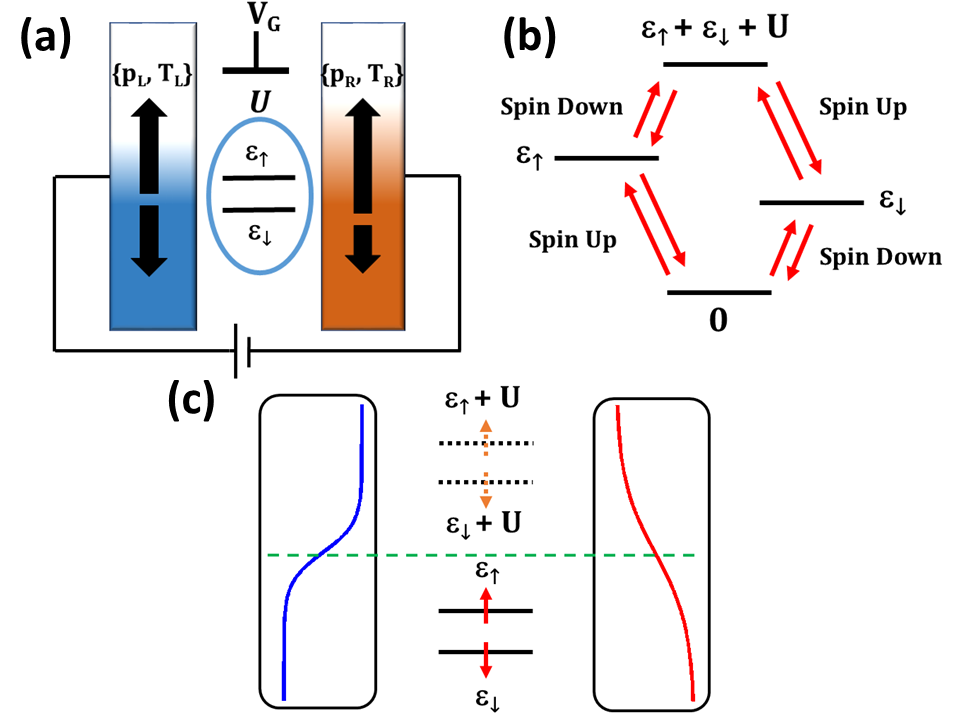}
	\caption{(a) Generic quantum dot thermoelectric setup with spin-split levels $\epsilon_{\uparrow}$ and $\epsilon_{\downarrow}$ which are weakly coupled to the contacts that may in general be ferromagnetic in the collinear configuration. The battery indicates a possible charge voltage bias. The right contact is at a hotter temperature. A gate voltage $V_G$ directly translates the energy spectrum with a magnitude proportional to its electrostatic coupling to the setup. (b) Fock space picture of the system in (a), with four occupation levels. The annotations 'Spin Up/Down' indicate the type of electron spin represented by the transition between the relevant states. (c) The single-particle representation of (b), with two transport channels $\epsilon_{\uparrow/\downarrow}$ and their Coulomb charged `partner' channels. The source drain bias is 0 and energies are defined with the reference $\mu_L = \mu_R = \mu = 0$. The electronic distribution in the contacts is described by the Fermi-Dirac distribution at their respective temperatures, which is schematically depicted as the blue curve in the left contact and the red curve in the right contact respectively.}
\label{Setup}
\end{figure}
\indent While the topic of thermoelectric pure spin current generation has been addressed in different kinds of low dimensional systems \cite{vanWees,Sanchez}, zero-dimensional systems such as magnetic molecules \cite{Timm_2,Wang} and quantum dots \cite{Leo_review,Basky} provide the most approachable test beds to appreciate the related concepts theoretically. Therefore, in recent works, the possibility of using a quantum dot or a magnetic molecule as a thermoelectric pure spin current generator or equivalently, a heat-to-pure-spin-current (HPSC) converter has generated some interest\cite{DiVentra,Ramsak,Wang}. While the  generation of spin polarized currents is a topic of immense interest, and is specifically carried out in the linear response regime \cite{Barnas_1,Zhang_1,Zheng_ZT,Jansen}, the generation of pure spin currents involves the cancellation of charge transport as an extra constraint.\\
\indent It is well known that an HPSC converter may be realized using a non-interacting quantum dot with a spin-split energy spectrum under particle hole symmetry conditions \cite{Ramsak}. However, the conditions under which an optimal HPSC converter may be realized is not well understood. For example, it is expected that Coulomb interaction, ubiquitous in typical quantum dot systems, may lead to non-trivial physics. Furthermore, the role of contact polarization in the setup is also not very obvious. The objective of this work is to hence provide a detailed picture of pure spin currents at various Coulomb interaction parameters $U$ and describe the optimal conditions that yield the largest terminal pure spin currents.\\
\indent The starting point to visualize a thermoelectric pure spin current is via a spin split energy level connected to the contacts, which may in general be magnetic, as shown in Fig.~\ref{Setup}(a). The spin-splitting may be realized via an external magnetic field in the case of a non-magnetic quantum dot \cite{Ramsak}, and via internal anisotropy fields in the case of a magnetic molecule \cite{Wang}.  In the non-interacting case, when the setup is gate tuned to particle-hole symmetry, an equal and opposite flow of the up and down spin currents result. This leads to a zero charge current and hence a pure spin current, as already noted in earlier works \cite{Ramsak,DiVentra,Wang}. In the presence of Coulomb interaction $U$,  the four transitions in the Fock space shown in Fig.~\ref{Setup}(b), result in additional transport channels as depicted in Fig.~\ref{Setup}(c). The presence of these additional correlated transport channels will introduce typical non-trivialities as a result of Coulomb interaction, to be discussed in detail here.\\
\indent In our detailed consideration of the above setup, we first demonstrate that an optimal performance is obtained when the contacts are non-magnetic. Turning to the role of the Coulomb interaction $U$, a non-trivial aspect pointed out here is that at sufficiently large values of $U (\ge U_0)$, pure spin currents tend to optimize at points other than where the particle-hole symmetry occurs. The fact that there exist pure-spin current points other than the point of particle-hole symmetry is corroborated by analyzing the conditions for an exact cancellation of charge transport between the energy levels $\epsilon$ and their Coulomb-charged partner levels $\epsilon+U$. It is also ascertained that a global maximum of pure spin current is generated at a typical value of the interaction parameter $U$. The optimal conditions derived in this paper may be realized using a typical gated quantum dot transport setup \cite{Tarucha_QD,Basky}.
\section{setup and formulation}
Our setup consists of an interacting single quantum dot coupled weakly to contacts which are maintained at temperatures $T_L$ and $T_R$ with $T_R > T_L$ as depicted in 
Fig.~\ref{Setup}(a). The contacts may be ferromagnetically polarized also, in which case, we assume a collinear configuration. The theoretical description of transport in our setup begins by defining the overall Hamiltonian $\hat{H}$ 
which is usually written as $\hat{H}=\hat{H}_D + \hat{H}_{C} + \hat{H}_{T}$, where $\hat{H}_D, \hat{H}_C$ and $\hat{H}_{T}$ represent the dot, 
reservoir and reservoir-dot coupling Hamiltonians respectively. In this paper, we use a non-magnetic quantum dot modeled as a single orbital Anderson impurity described by Hamiltonian:
\begin{equation}
\hat{H}_D=\sum_{\sigma} \epsilon_{\sigma} \hat{n}_{\sigma} + U \hat{n}_{\uparrow} \hat{n}_{\downarrow},
\label{Ham_def}
\end{equation}
where $\epsilon_{\sigma}$ represents the orbital energy, $\hat{n}_{\sigma} = \hat{d}_{\sigma}^{\dagger} \hat{d}_{\sigma}$ is the occupation number operator of an electron with 
spin $\sigma = \uparrow$, or $\sigma = \downarrow$, and $U$ is the Coulomb interaction energy between electrons of opposite spins occupying the same orbital. The resulting four Fock-space states are $|0 \rangle, \mid \uparrow \rangle, \mid \downarrow \rangle$ and $\mid \uparrow \downarrow \rangle$, labeled by their many-electron eigen-energies $0$, $\epsilon_{\uparrow}$, $\epsilon_{\downarrow}$, and $\epsilon_{\uparrow}+\epsilon_{\downarrow} +U$ respectively.  Specific to our setup, based on the Fock space picture schematically shown in Fig.~\ref{Setup}(b), the degeneracy between the single electron levels $\epsilon_{\uparrow}$ and $\epsilon_{\downarrow}$ may be broken by the application of an external magnetic field or due to the presence of internal anisotropy fields, such as, in the case of magnetic molecules \cite{Timm_2}.
The contact Hamiltonian is given by 
$\hat{H}_{C} = \sum_{\alpha=L,R}\sum_{k \sigma_{\alpha}} \epsilon_{\alpha k \sigma_{\alpha}} \hat{n}_{\alpha k \sigma_{\alpha}}$, 
where $\alpha$ labels the left/right reservoir ($L$ or $R$ in our case) and the summation is taken over the single particle states labeled 
$\{k \sigma_{\alpha}\}$, and $\sigma_{\alpha}=\pm$ denotes the majority and minority spin orientation in the contacts. 
The tunneling Hamiltonian that represents the dot-contact coupling may in general be written as:
\begin{equation} 
\hat{H}_{T}=\sum_{\alpha k \sigma_{\alpha}} \left ( t_{\alpha} \hat{c}^{\dagger}_{\alpha k \sigma_{\alpha}} \hat{d}_{\sigma_{\alpha}} + 
{{t}^{\ast}_{\alpha}} \hat{d}^{\dagger}_{\sigma_{\alpha}} 
\hat{c}_{\alpha k \sigma_{\alpha}} \right ) 
\label{eq:Tunn_Ham}
\end{equation}
where $(\hat{c}^{\dagger},\hat{c})$ and $(\hat{d}^{\dagger},\hat{d})$ are the creation/annihilation operators of the reservoir states labeled $\{k \sigma_{\alpha} \}$ and of the quantum dot one particle states respectively, and $t_{\alpha}$ denotes the tunneling matrix element. At energies close to the Fermi level, metallic contacts 
can be described using a constant density of states, parameterized using 
the bare-electron tunneling energies $\gamma_{\alpha}=\sum_{k \sigma} {2 \pi}|t_{\alpha k \sigma, s}|^2 \delta(E-\epsilon_{k \sigma})$, with  $(\alpha=L/R)$ representing the left or right contact. In the case of ferromagnetic contacts, one often uses a Stoner model \cite{Milena_noncoll,Basky_Milena} to obtain {\it{spin resolved}} tunneling energies for the majority (minority) bands, or equivalently the $\uparrow (\downarrow)$ bands as $\gamma^{\sigma}_{\alpha}=\sum_{k} {2 \pi}|t_{\alpha k \sigma, s}|^2 \delta(E-\epsilon_{k \sigma})$, where $\sigma=\uparrow (\downarrow)$. The contacts are then characterized by lead polarizations $p_\alpha$, defined as
\begin{equation}
p_\alpha = \frac{\gamma_{\alpha}^{\uparrow} - \gamma_{\alpha}^{\downarrow}}{\gamma_{\alpha}^{\uparrow}+\gamma_{\alpha}^{\downarrow}}.
\end{equation}
\subsection{Charge and Spin currents}
Most quantum dot experiments \cite{Tarucha_QD,Leo_review} are performed in the weakly coupled sequential tunneling limit. Hence, in this limit, i.e., when the energy scale associated with the bare electron tunneling rate $\gamma_{\alpha}^\sigma$ of either contact $\alpha=L/R$ is much smaller than the ambient temperature ($\gamma_{\alpha}^\sigma << k_BT$), the transport is described by the master equation approach \cite{Beenakker,Basky_Beenakker,Basky_Datta,Timm}, and more generally by the density matrix approach \cite{Koenig_1,Koenig_2,Koenig_3,Brouw,Basky_Milena} in the Fock space of the quantum dot Hamiltonian. \\
\indent Specific to our setup, due to the collinear configuration considered \cite{Timm,Koenig_1}, one solves a set of master equations for the non-equilibrium probabilities $P_1,P_2,P_3$ and $P_4$ of the four Fock space states $|0 \rangle, \mid \uparrow \rangle, \mid \downarrow \rangle$ and $\mid \uparrow \downarrow \rangle$ given by:
\begin{equation}
\frac{dP_{i}}{dt} = \sum_{j}(-R_{{i}\rightarrow {j}}P_{i} +  R_{{ j}\rightarrow {i}}P_{j}) ,
\label{ebeenakker}
\end{equation}
where the index $j$ runs over states that encompass the possible transitions from the state labeled by the index $i$. The tunneling transition probabilities or simply transition rates are $R_{{i} \rightarrow {j}}$ between states $|i \rangle$ and $|j \rangle$ that differ by an electron number such that $R_{i \rightarrow j} =\sum_{\alpha=L,R} \frac{\gamma^{\sigma}_{\alpha}}{\hbar} f \left( \frac{\epsilon^{a\sigma}_{ij}-\mu_{\alpha}}{k_BT_{\alpha}} \right )$ for transitions representing the addition of an electron with spin $\sigma=\uparrow(\downarrow)$, and $R_{i \rightarrow j} =\sum_{\alpha=L,R} \frac{\gamma^{\sigma}_{\alpha}}{\hbar} \left[ 1-f \left( \frac{\epsilon^{r\sigma}_{ij}-\mu_{\alpha}}{k_BT_{\alpha}} \right ) \right ]$ for transitions representing the removal of an electron with spin $\sigma=\uparrow(\downarrow)$. The contact electrochemical potentials and temperatures are respectively labeled as $\mu_{\alpha}$ and $T_{\alpha}$, and $f$ is the corresponding Fermi-Dirac distribution function with the single particle removal and addition
transport channels for a given spin $\sigma=\uparrow (\downarrow)$, given by  $\epsilon^{r\sigma}_{ij} = E_i - E_j$ and $\epsilon^{a\sigma}_{ij} =E_j - E_i$ respectively. \\    
\indent Applying the formalism described above, the charge (spin) current through the dot is given by the sum (difference) of the currents carried by the up-spin ($\sigma=\uparrow$) channels and the down-spin ($\sigma=\downarrow$) channels, which are individually depicted in Fig.~\ref{Setup}(c). The steady-state up spin (down spin) current associated with either contact may be formulated by setting $\frac{dP_i}{dt}=0$ in \eqref{ebeenakker}, to obtain the steady state probabilities $P_i$, following which we have
\begin{equation} 
I^{\alpha}_{\uparrow (\downarrow)} = -q \sum_{i} \sum_{j>i}(R^{\alpha \uparrow (\downarrow)}_{{i}\rightarrow {j}}P_{i} - R^{\alpha \uparrow (\downarrow)}_{{ j}\rightarrow {i}}P_{j}),
\label{eq:curr}
\end{equation}
with $P_i$'s representing the steady state probabilities obtained, and and $q$ being the electronic charge. The transition rate $R^{\alpha \uparrow (\downarrow)}_{{i}\rightarrow {j}}$ represents an allowed up spin (down spin) transition associated with contact $\alpha=L/R$ and the summation runs over all the possible addition (removal) transitions dictated in the Fock space. Specific to our setup, the net up-spin current is given by the sum over the transitions represented by the transport channels $\epsilon_\uparrow$ and $\epsilon_\uparrow + U$, while the net down-spin current corresponds to that represented by the transport channels $\epsilon_\downarrow$ and $\epsilon_\downarrow + U$. 
\begin{figure}
\centering
	\includegraphics[width=2.85in, height = 2.3in]{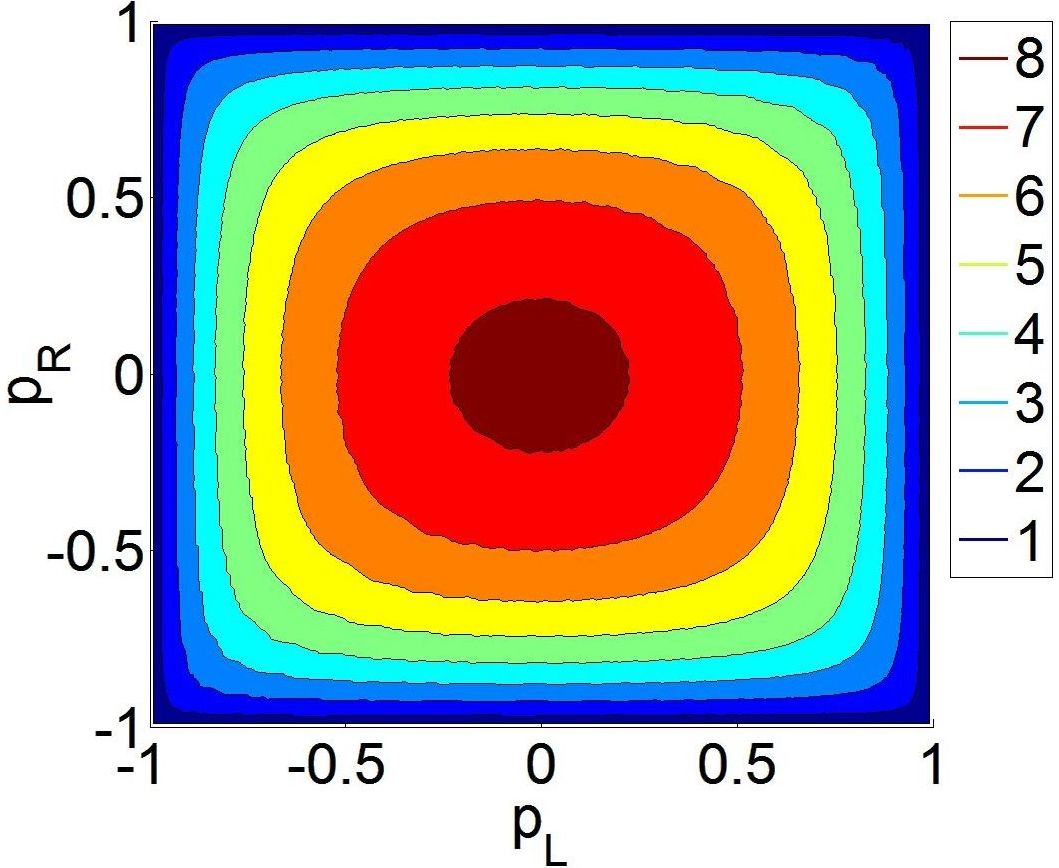}
	\caption{Contour plot of pure spin current produced at lead polarizations $p_L$ and $p_R$, where the spin-dependent strength of contact coupling is $\gamma(1\pm p_{L/R})$. An external magnetic field $B = 1$ T is applied. The legend depicts the spin current in picoamperes (pA) at each contour. It is seen that the pure spin current is maximum for $p_L = p_R = 0$, i.e. for non-magnetic contacts.}
\label{Contour}
\end{figure}
\section{Optimizing pure spin currents}
We first attempt to understand the role of contact polarizations in optimizing the magnitude of the terminal spin currents. One may compute charge and spin currents using \eqref{eq:curr} for various values of lead polarizations $p_\alpha$. Analyzing electron transport using the above formalism, it is seen that, while spin polarized current magnitudes in general are larger when the quantum dot is coupled with ferromagnetic contacts, the value of the \textit{pure} spin current diminishes with increasing contact polarization. In Fig.~\ref{Contour}, we depict the pure spin current magnitudes as a contour plot for $U = 1.5k_BT$ for a complete range of contact polarizations $-1 \le p_L, p_R \le 1$. It is seen that pure spin current is largest for $p_L = p_R = 0$. This is because at larger contact polarizations, charge transport becomes increasingly spin polarized, to the point where charge and spin currents are identical when $p_L = p_R = \pm 1$, giving rise to a highly spin polarized current but no pure spin current. Since our intention is to maximize pure spin currents, we shall henceforth consider only non-magnetic contacts such that $\gamma_{\alpha}^{\uparrow} = \gamma_{\alpha}^{\downarrow} = \gamma := (\gamma_{\alpha}^{\uparrow} + \gamma_{\alpha}^{\downarrow})/2$ for $\alpha = L,R$. We set $\gamma = 5 \ \mu$eV here. \\

\indent Using \eqref{eq:curr}, and defining $f\equiv f_L - f_R$, it can be shown that the steady-state up and down spin channel currents $I_{\uparrow/\downarrow}$ are given by
\begin{eqnarray}
I_\uparrow &=& -\frac{q\gamma}{\hbar}\{f(\epsilon_\uparrow)(P_1 + P_2) + f(\epsilon_\uparrow + U)(P_3 + P_4)\} \label{iup} \\
I_\downarrow &=& -\frac{q\gamma}{\hbar}\{f(\epsilon_\downarrow)(P_1 + P_3) + f(\epsilon_\downarrow + U)(P_2 + P_4)\}.
\label{idn}
\end{eqnarray}
The charge current may be calculated as $I = I_\uparrow + I_\downarrow$ and the spin current as $I_s = I_\uparrow - I_\downarrow$. A pure spin current is generated in the quantum dot when the up-spin and down-spin currents are equal in magnitude but opposite in sign. It was pointed out in earlier works \cite{DiVentra,Ramsak}, that this setup may be tuned to generate pure spin currents by applying an appropriate gate potential $V_G$ as shown in the schematic of Fig.~\ref{Setup}(a). \\

\begin{figure}
\centering
	\includegraphics[width=3.1in, height = 2.5in]{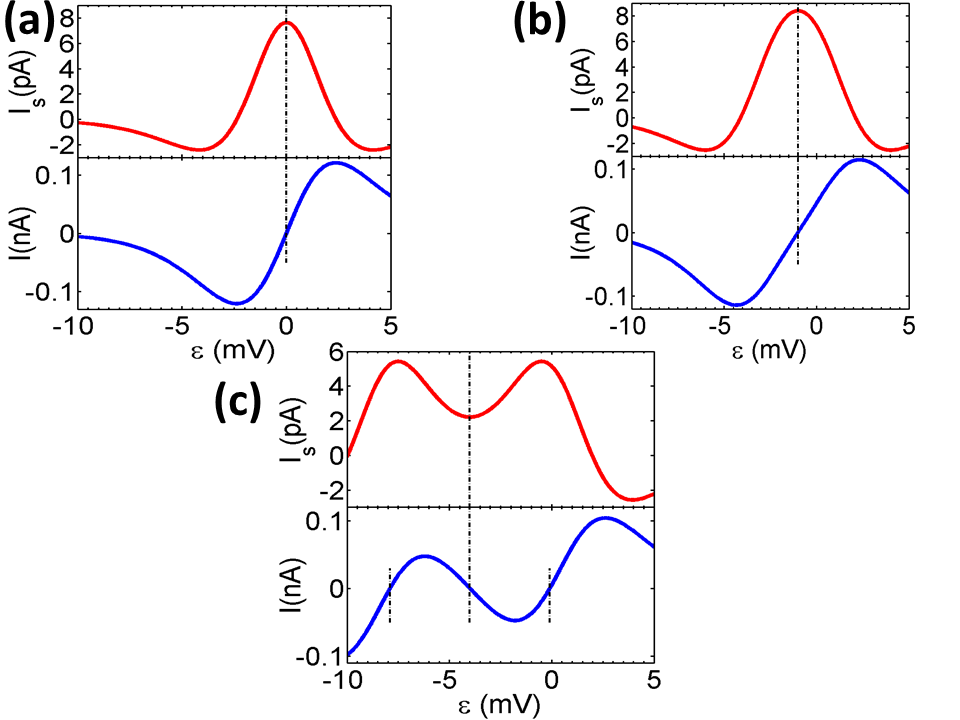}
	\caption{Spin currents versus gate potential energy. (a) and (b) The top panel depicts the spin current and the bottom panel, the charge current for $U = 0$ and $U = 3$ meV respectively. Note that since these values of $U$ are less than $2.4k_BT$, there is only one point where charge current is zero and hence the spin current is maximum. (c) The same plots for $U = 8$ meV $> 2.4k_BT$. We can see three optima in the spin current panel: one at $\epsilon = -U/2$ (a minimum) and two other points where the spin current is maximum. Note the transition from maximum to minimum spin current at $\epsilon = -U/2$ from cases (a-b) to (c). Since we are interested in pure spin currents, these are obtained at $\epsilon = -U/2$ and $\epsilon = -U/2 \pm D$, marked by dotted black lines in the figure.}
\label{Currents}
\end{figure}

\indent It was also pointed out that for relevant contact temperatures $T_{L/R}$, a pure spin current is generated at the particle-hole symmetry point \cite{Ramsak} corresponding to a gate potential energy of $-U/2$, where $U$ is the Coulomb charging energy. This may be understood as follows:  At $\epsilon = -U/2$, we have the two energy levels at $\epsilon_{\uparrow/\downarrow}=-U/2 \pm \delta$, where $\delta = \frac{1}{2}g\mu_BB$. Their corresponding `Coloumb-charged' partner levels are located at $\epsilon_{\uparrow/\downarrow} + U = U/2 \pm \delta$ (see Fig.~\ref{Setup}(c)). When the spin down level is conducting at $-U/2 - \delta$, its Coulomb charged partner (spin up) level conducts at $U/2 + \delta$. Since these two levels are located symmetrically about $\mu = 0$, they conduct equal amounts of current in opposite directions. Similarly, when the spin up level conducts at $-U/2 + \delta$, its Coulomb charged partner (spin down) level conducts at $U/2 - \delta$, which are symmetric about $\mu = 0$ as well. Therefore, the net charge current conducted at $\epsilon = -U/2$ is zero regardless of the value of $U$. \\
\indent In the discussion to follow, we will first show that while a pure-spin current is indeed generated at the particle-hole symmetry point, this may not be the only point if $U$ is sufficiently large. Further, we show that the magnitude of the spin current generated at the particle-hole symmetric point in such a case would be smaller than that of the other points where the pure-spin current is generated. We then discuss the relationship between $U$ and the average contact temperature for such a condition to arise.\\
\indent In Figs.~\ref{Currents}(a), ~\ref{Currents}(b) and ~\ref{Currents}(c), we depict the charge (blue) and spin (red) currents versus gate potential energy $\epsilon$ for various values of $U$ in the absence of an applied bias, i.e., $\mu_L = \mu_R = 0$. We assume the spin accumulation to be negligible here, which would be the case when the non-magnetic contacts have a conductivity much higher than the linear response conductivity of the quantum dot. The plots in Fig.~\ref{Currents}(a) correspond to $U = 0$. The first straightforward result is that in the non-interacting case, pure spin currents are generated at $\epsilon = -U/2 = 0$, because the two levels $\epsilon_\uparrow = \delta = -\epsilon_\downarrow$ conduct equal currents in opposite directions simultaneously leading to zero charge current.\\
\indent Since any realistic quantum dot transport setup \cite{Tarucha_QD,Leo_review} has a non-zero and finite $U$, we shall focus on the interacting setup in the rest of the paper. Consider the scenario depicted in Fig.~\ref{Currents}(b) for $U = 3$ meV, where it is seen that the particle-hole symmetry point of $\epsilon = -U/2$ yields a zero charge current and a pure spin current. Here, the temperature gradient is set based on the Carnot efficiency of $\eta = 20\%$ \cite{Basky}, such that $T_R = 20$ K and consequently $T_L = (1-\eta)T_R = 16$ K. For a better analytical treatment, the spin current $I_{s}$ may be simplified in the case of $\epsilon = -U/2$ using the relation $\epsilon_\uparrow + \epsilon_\downarrow + U = 0$ to yield
\begin{equation}
I_s(-\frac{U}{2}) = -\frac{q\gamma}{\hbar}\{f(\epsilon_\uparrow) - f(\epsilon_\downarrow) + \{f(\epsilon_\uparrow) + f(\epsilon_\downarrow)\}(P_3 - P_2)\}
\label{isspec}
\end{equation}
\begin{figure}
\centering
	\includegraphics[width=3.1in, height = 2.5in]{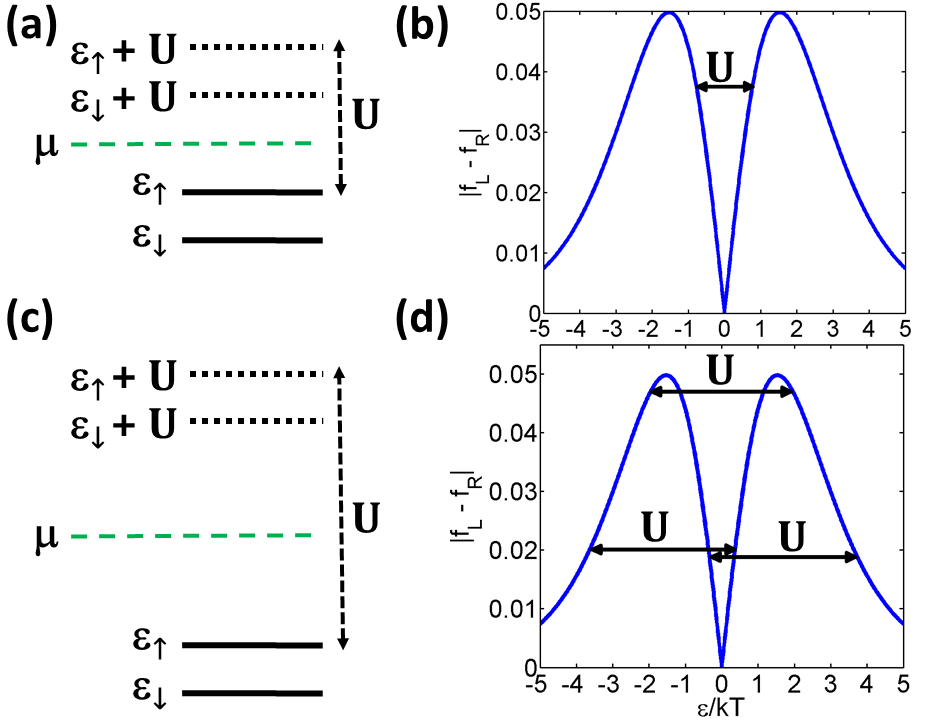}
	\caption{Heuristic explanation for the spin current behavior in Fig.~\ref{Currents} (b-c). Current through a level positioned at $x$ is approximately modeled by $(f_L - f_R)(x)$. In the one particle picture, the difference in energy between the mid-points $\epsilon$ and $\epsilon+U$ equals $U$, which must correspond to the points with equal values of $f_L-f_R$ in order to cancel the charge current. (a) Single particle picture for small $U$, and (b) a plot of $|f_L-f_R|$ depicting the only possible configuration where currents through the levels below $\mu$ and above $\mu$ are equal in magnitude and opposite in sign, i.e., at $\epsilon = -U/2$. (c) The single particle picture for larger $U$, and (d) a plot of $|f_L - f_R|$ showing that there are three possible configurations where points separated by $U$ have equal values of the aforementioned function. Note that the set of points other than $\epsilon = -U/2$ lie along the same straight line due to the symmetry of $|f_L-f_R|$. However, the double-arrows have been shown slightly skewed to indicate the presence of two distinct values of $\epsilon$ where the charge current has zero magnitude.} 
\label{Fermi}
\end{figure}
where $f \equiv f_L - f_R$. At $\epsilon = -U/2$, the probabilities may be analytically obtained by solving the rate equations, given the symmetry in our case. By expanding the Fermi-Dirac distributions in \eqref{isspec} using the Taylor approximation, given $g\mu_BB \ll k_BT$ with $T = (T_L + T_R)/2$ and $\Delta T = T_L - T_R$, it can be shown (see Appendix) that 
\begin{equation}
I_s(-\frac{U}{2}) \approx -\frac{q\gamma}{\hbar}\cdot\frac{2\delta\Delta T}{T}\cdot\frac{f_0(1-f_0)}{k_BT}\cdot\{1+\frac{U}{2k_BT} -\frac{U}{2k_BT}f_0\}, 
\label{i_approx}
\end{equation} 
where $\delta = g\mu_B B$ and $f_0$ is the Fermi-Dirac distribution evaluated at $T$ and $\epsilon = -U/2$. Several interesting points may be made from this analysis. To begin with, one may show that if $U$ is sufficiently large, the pure spin current is not a maximum at the point of particle-hole symmetry, i.e., at $\epsilon = -U/2$. Indeed, there exist two other points where pure spin currents which are larger in magnitude than that at $\epsilon = -U/2$. This scenario is depicted in Fig.~\ref{Currents}(c) for $U = 8$ meV, $T_L = 16$ K and $T_R = 20$ K. Using elementary perturbation techniques to the spin current at $\epsilon = -U/2$ from \eqref{i_approx}, by computing its inflection point in $\epsilon$, one can show that such a transition (for small $\eta$) from having one zero of the charge current (such as in Fig.~\ref{Currents}(b)) to having three zeros of the charge current (Fig.~\ref{Currents}(c)) occurs at 
\begin{equation}
U_0 \approx 2.6k_BT
\end{equation}
Therefore, the crucial point to note here is that the optimal gate potential energy that generates the maximum pure spin current equals the negative half of Coulomb charging energy, only when $U \le U_0$ for relevant contact temperatures. \\
\subsection{The optimal points}
\begin{figure}
\centering
	\includegraphics[width=3.28in, height = 1.64in]{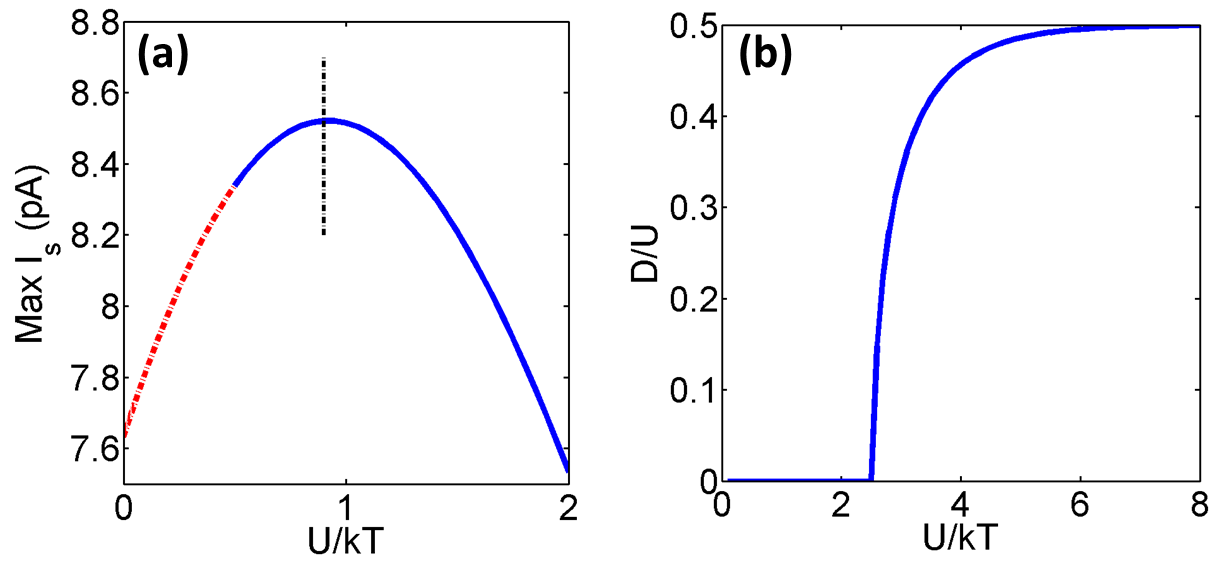}
	\caption{(a) Maximum pure spin current for various values of $U$. The optimal choice of $U$ is denoted by the black line. The dotted red line for low $U$ indicates the region where sequential tunneling approximation may be invalid, specifically when $\gamma_\alpha \approx U, k_BT$. (b) Plot of $D/U$ vs. $U/k_BT$, where the additional zeros of charge current occur at $\epsilon = -U/2 \pm D$. Note that $D\rightarrow U/2$ for large $U$.}
\label{Optima}
\end{figure}

In order to understand this transition in behavior across $U = U_0$, we turn to Fig.~\ref{Fermi}. Heuristically, for $\mu_L = \mu_R$, electron transport through a level positioned at $x$ is given by the difference $|(f_L - f_R)(x)|$, with the sign depending on the location of $x$ with respect to the electrochemical potential $\mu_{L/R}$. For purposes of computing charge currents, instead of considering the spin splitting of the levels $\epsilon_{\uparrow/\downarrow}$ and their Coulomb-charged partner levels $\epsilon_{\uparrow/\downarrow}+U$, we could analyze the mid points $\epsilon$ and $\epsilon+U$. These two levels are separated by $U$ along the $\epsilon$-axis. This scenario is depicted in Figs.~\ref{Fermi}(a) and ~\ref{Fermi}(c), with their plots of $|f_L-f_R|$ in Figs.~\ref{Fermi}(b) and ~\ref{Fermi}(d), respectively. Note that $|f_L-f_R|$ is symmetric about $\epsilon = 0$ since $\mu_L = \mu_R = 0$. In order to cancel out charge currents, we need a point on the $\epsilon$-axis to the left of $\epsilon$ = 0, and one to the right of $\epsilon =0$ (Figs.~\ref{Fermi}(b,d)), since these conduct in opposite directions. When $U$ is sufficiently small, it is seen (Figs.~\ref{Fermi}(a-b)) that there is only one possible configuration where two points on the opposite sides of $\epsilon = 0$, separated by a distance $U$, have an equal value of $|f_L-f_R|$. This point is $\epsilon = -U/2$ (and $\epsilon+U = U/2 = -\epsilon$), since $|f_L-f_R|$ is symmetric. However, for large values of $U$ (Fig.~\ref{Fermi}(c-d)), there are two other points in addition to $\epsilon = -U/2$ where points separated by a distance $U$ have equal values of $|f_L-f_R|$. Approximately, one would expect this change of behavior to occur when $U$ crosses over the maxima of $|f_L-f_R|$ (see Fig.~\ref{Fermi}(b,d)). Indeed, the distance between the maxima of $|f_L-f_R|$ is about $3k_BT$, while a more detailed calculation gives this transition point as $U_0 = 2.6k_BT$. The proximity of these two values justifies the heuristic explanation considered here.\\
\indent Let us first consider the case of $U \le U_0$. In this scenario, there is a single zero of charge current which is simultaneously the maximum of pure spin current. However, the pure spin current is not equal for all values of $U \le U_0$. To obtain the maximal pure spin current at this point of particle-hole symmetry, one can maximize $I_s(-U/2)$ (see Eq. \eqref{i_approx}), such that for small $\eta$, the maximum pure spin current is generated at 
\begin{equation} 
U_m \approx 0.92k_BT
\end{equation}
A plot of the pure spin-current at the particle hole symmetry point versus $U/k_BT$ is shown in Fig.~\ref{Optima}(a), depicting this discussion. 
\\ \indent On the other hand, when $U \ge U_0$, the charge current is zero at three points, i.e., $\epsilon = -U/2$ and $\epsilon = -U/2 \pm D$ (the two other points are equidistant from $\epsilon = -U/2$ due to the symmetry of transport across $\mu_{L/R} = 0$), where $D/U$ is plotted in Fig.~\ref{Optima}(a) as a function of $U/k_BT$. An analytical approximation to $D$ can be obtained as
\begin{equation*}
D \approx \sqrt{0.24(U-U_0)^2 + 1.4(U-U_0)k_BT}, \ \ U \ge U_0.
\end{equation*}
Therefore, the other two zeros of the charge current arise only for $U\ge U_0$. For $U \gg U_0$, it can be seen that $D \approx \sqrt{0.24(U-U_0)^2} \approx \sqrt{0.24U^2} \approx U/2$, which means that charge current goes to zero at $\epsilon = 0$ and $-U$ in addition to $\epsilon = -U/2$. While pure spin currents may be tapped at any of $\epsilon = -U/2$ and $\epsilon = -U/2 \pm D$, it is clear from Fig.~\ref{Currents}(c) that the pure spin currents at the latter two gate energy points are higher. Thus, when $U \ge U_0$, pure spin currents are maximized at points other than where the particle-hole symmetry occurs.\\
\indent We have thus described the optimization of pure spin currents for different values of the Coulomb interaction parameter $U$ by making appropriate choices for the gate potential energy $\epsilon$. We discussed the transition in behavior of the charge and spin currents around $U = U_0$ and its impact on our choice of gate potential energy to produce optimal pure spin currents. For $U \le U_0$, one ought to bias the system at the particle-hole symmetric point $\epsilon = -U/2$, while for $U \ge U_0$, the optimal choice is to bias the system at $\epsilon = -U/2 \pm D$. In particular, the point $\epsilon = -U/2$ becomes a local minimum for the pure spin current for $U > U_0$.
\section{Discussion}
Our analysis above has been based on the evaluation of non-equilibrium charge and spin currents without any mention of thermodynamic efficiency. In the field of charge based thermoelectrics, it is often the case that a linear response analysis \cite{Basky,goldsmid} is used as a substitute for actual non-equilibrium transport calculations. In such an analysis, one analyzes the linear response parameters related to the electric and heat currents. It is then customary to define a figure of merit $zT$ \cite{goldsmid} that directly relates to the maximum efficiency by using the linear response parameters as $zT=\frac{S^2 \sigma T}{\kappa}$, where $S$, $\sigma$, $\kappa$ and $T$ are the Seebeck coefficient, electronic conductivity, thermal conductivity and the operating temperature respectively. This is based on the calculation of maximum efficiency of a simple circuit \cite{goldsmid} consisting of the thermoelectric device and a resistive load. It is thus tempting to draw an inspired analogy in the case of spin based thermoelectrics \cite{DiVentra,Zheng_ZT} and define a spin figure of merit $z_sT = \frac{S_s^2|\sigma_s|T}{\kappa}$, where $S_s$ is the spin-Seebeck coefficient and $\sigma_s$ is the spin conductivity. \\
\indent While such an analogy might work in a case where the spin polarized current eventually drives a charge current based load \cite{Bauer_ZT}, reproducing the derivation for such a figure of merit in the case of a pure spin circuit is complicated by two factors. One, since the spin current decays within the metallic contact, an additional resistance plays a role in finding the efficiency. Two, the `load' in a spin circuit is often a magnetic logic device, for instance, a nanomagnet whose magnetization direction needs to be flipped or maintained at a particular micromagnetic orbit. This cannot be approximated by a resistor; in other words, the energy consumed by such a device is often a complicated function of the spin current supplied as well as the trajectory taken by the magnetization vector. Therefore, a $z_sT$ derived from a spin circuit modelled as a spin thermoelectric device with a spin-resistor (analogous  to the charge circuits) is unlikely to be sufficient to be a figure-of-merit of any real setup.
\section{Conclusion} In this paper, we have analyzed a quantum dot based heat-to-pure-spin-current converter in relation to the generation of maximal terminal pure spin currents. We provided a detailed picture of thermoelectric pure spin currents at various Coulomb interaction parameters $U$ and described the conditions necessary for an exact cancellation of charge transport between energy levels, so as to yield the largest terminal pure spin currents. A non-trivial aspect pointed out here was that at sufficiently large values of $U$ ($\ge U_0$), pure spin currents tend to optimize at points other than where the particle-hole symmetry occurs. It was also ascertained that a global maximum of pure spin current is generated at a typical value of the interaction parameter $U$. These optimum conditions may be easily realized using a typical gated quantum dot thermoelectric transport setup \cite{Tarucha_QD,Leo_review}. Apart from demonstrating numerous non-trivialities with respect to pure spin current optimization, we have also attempted to connect with the linear response picture. We believe that making a connection with conventional linear response wisdom inspired from charge based thermoelectrics is a bit misguiding, specifically when the topic of utilizing pure spin currents is concerned. Such an analysis might make sense if one re-converts a spin polarized charge current back to charge current, such that the entire setup mimics the objective of a charge thermoelectric setup \cite{Bauer_ZT}. It would be an interesting venture to quantify a figure of merit specifically focussed on utilizing the pure spin current for performing a particular micromagnetic operation, such as the flipping of an ultra thin magnet. Whether such a figure of merit is possible and significant merits a detailed consideration of the composite setup comprising the quantum dot thermoelectric device, metallic leads and a micromagnetic spin-load.\\
{\it{Acknowledgments:}} This work was in part supported by the IIT Bombay SEED grant and the Department of Science and Technology (DST), India, under the Science and Engineering Board grant no. SERB/F/3370/2013-2014. We would also like to acknowledge the center of excellence in nanoelectronics. Useful discussions with Prof. Ashwin Tulapurkar are gratefully acknowledged. 
\appendix
\section{Approximation for spin currents at the electron-hole symmetry point}
We begin with 
\begin{equation}
I_s(-\frac{U}{2}) -\frac{q\gamma}{\hbar}\{f(\epsilon_\uparrow) - f(\epsilon_\downarrow) + \{f(\epsilon_\uparrow) + f(\epsilon_\downarrow)\}(P_3 - P_2)\}.
\label{iscopy}
\end{equation}
At $\epsilon = -U/2$, the probabilities assume analytical forms due to the symmetry in the energy levels about $\mu$. One may check that 
\begin{eqnarray*}
P_2 &=& \frac{F_\uparrow F_\downarrow - 2F_\uparrow}{F_\uparrow + F_\downarrow - 4} \\
P_3 &=& \frac{F_\uparrow F_\downarrow - 2F_\downarrow}{F_\uparrow + F_\downarrow - 4}, \\
\end{eqnarray*}
where $F_\sigma = f_L(\epsilon_\sigma) + f_R(\epsilon_\sigma)$, $f \equiv f_L - f_R$ and $\epsilon_\sigma = \epsilon +\sigma g\mu_BB = \epsilon + \sigma\delta$. Next, we expand
\begin{equation*}
f(\epsilon_\sigma) \approx f(\epsilon) + \sigma\delta\cdot\big(\frac{df}{d\epsilon}\big).
\end{equation*}
Since there is no applied bias and the levels $\epsilon$ are defined with $\mu = 0$ as the reference, we have 
\begin{equation}
f(\epsilon_\sigma) \approx 0 + \big(-\frac{df_0}{d\epsilon}\big)\frac{\epsilon}{T}\Delta T + \sigma\delta\cdot\frac{\Delta T}{T}\{\big(-\frac{df_0}{d\epsilon}\big) + \big(-\frac{d^2f_0}{d\epsilon^2}\big)\epsilon\}
\label{approx}
\end{equation}
using the first order derivative approximations. Here $f_0$ is simply the Fermi Dirac function evaluated at $\epsilon$.	The latter two terms are derived by substituting the approximation for $f$ used in the first two terms into $df/d\epsilon$. Since we have
\begin{eqnarray*}
-\frac{df_0}{d\epsilon} &=& \frac{f_0(1-f_0)}{k_BT} \\
\frac{d^2f_0}{d\epsilon^2} &=& \frac{f_0(1-f_0)(1-2f_0)}{(k_BT)^2},
\end{eqnarray*}
we use \eqref{approx} and the above expressions to get
\begin{eqnarray}
f(\epsilon_\uparrow) - f(\epsilon_\downarrow) &=& \frac{2\delta\Delta T}{T}\frac{f_0(1-f_0)}{k_BT}(1- \frac{1-2f_0}{k_BT}\epsilon) \label{app1} \\
f(\epsilon_\uparrow) + f(\epsilon_\downarrow) &=& \frac{\Delta T}{T}\frac{f_0(1-f_0)}{k_BT}\epsilon \label{app2}.
\end{eqnarray}
Finally, we obtain
\begin{equation*}
P_3 - P_2 = \frac{2(F_\uparrow - F_\downarrow)}{F_\uparrow + F_\downarrow - 4}.
\end{equation*}
To approximate $F_\sigma$, we note that
\begin{equation*}
F_\sigma = f_L(\epsilon_\sigma) + f_R(\epsilon_\sigma) = 2f_0 + 2\sigma\delta\cdot\frac{df_0}{d\epsilon},
\end{equation*}
where $f_0$ is directly used as an approximation since the temperature difference on the $L/R$ contacts cancel off upon adding their Fermi distributions. Therefore, 
\begin{equation}
P_3 - P_2 = \frac{8\delta(df_0/d\epsilon)}{4f_0 - 4} = -\frac{2\delta f_0(1-f_0)}{k_BT(f_0-1)} = \frac{2\delta f_0}{k_BT}.
\label{app3}
\end{equation}
Now, plugging Eqs. \eqref{app1}, \eqref{app2} and \eqref{app3} into \eqref{iscopy}, and replacing $\epsilon$ with $-U/2$,  we obtain 
\vskip 0.1in
\begin{eqnarray}
I_s &=& -\frac{2\delta\Delta T}{T}\{\frac{f_0(1-f_0)}{k_BT}(1+ \frac{(1-2f_0)U}{2k_BT}) + \frac{f_0^2(1-f_0)U}{2(k_BT)^2}\} \nonumber \\ 
&=& -\frac{q\gamma}{\hbar}\cdot\frac{2\delta\Delta T}{T}\cdot\frac{f_0(1-f_0)}{k_BT}\cdot\{1 + \frac{U}{2k_BT}(1 -f_0)\},
\end{eqnarray}
which is the desired expression. It should be noted that $f_0 \equiv f_0(-\frac{U}{2})$ at this point. Therefore, the entire expression is a function of the reduced parameter $U/k_BT$. One may then perform maximization or calculation of inflection point for this expression to obtain $U_0$ and $U_m$ as described in the main text.
\bibliographystyle{apsrev}	
\bibliography{bibliography}

\end{document}